\documentclass[floatfix,aps,prd,showpacs,amsmath,nofootinbib,preprintnumbers,
twocolumn]
{revtex4}
\usepackage{graphicx}
\usepackage{colordvi}
    \def\be{\begin{equation}}
    \def\ee{\end{equation}}
    \def\ba{\begin{eqnarray}}
    \def\ea{\end{eqnarray}}

\begin{document}

\title{Stochastic growth of quantum fluctuations during slow-roll inflation}

\author{F. Finelli$\,^{1,2}$, G. Marozzi$\,^{3}$, A. A.
Starobinsky$\,^{4}$,
G. P. Vacca$\,^{5,2}$ and G. Venturi$\,^{5,2}$}

%\author{F. Finelli} \email{finelli@iasfbo.inaf.it}
\affiliation{$^{1}$ INAF/IASF Bologna,
Istituto di Astrofisica Spaziale e Fisica
Cosmica di Bologna \\
via Gobetti 101, I-40129 Bologna - Italy}
\affiliation{$^{2}$ INFN, Sezione di Bologna,
Via Irnerio 46, I-40126 Bologna, Italy}
%\author{G. Marozzi} \email{marozzi@bo.infn.it}
\affiliation{
%$^{3}$ ${\cal G}\mathbb{R}\varepsilon\mathbb{E}{\cal O}$ --
$^{3}$ GR$\varepsilon$CO --
Institut d'Astrophysique de Paris, UMR7095,
 CNRS,  \\
 Universit\'e Pierre \& Marie Curie, 98 bis boulevard Arago,
75014 Paris, France}
\affiliation{$^{4}$ Landau Institute for Theoretical Physics, Moscow, 119334,
Russia}
\affiliation{$^{5}$ Dipartimento di Fisica, Universit\`a degli Studi di
Bologna, via Irnerio, 46 -- I-40126 Bologna -- Italy}

%%%%%%%%%%%%%%%%%%%%%%%%%%%%%%%%%%%%%%%%%%%%%%%%%
\begin{abstract}
We compute the growth of the mean square of quantum fluctuations
of test fields with small effective mass during a slowly changing,
nearly de Sitter stage which takes place in different inflationary
models. We consider a minimally coupled scalar with a small mass,
a modulus with an effective mass $ \propto H^2$ (with $H$ the
Hubble parameter) and a massless non-minimally coupled scalar in
the test field approximation and compare the growth of their
relative mean square with the one of gauge invariant
inflaton fluctuations. We
find that in most of the single fields inflationary models the
mean square gauge invariant
inflaton fluctuation grows {\em faster} than any test
field with a non-negative effective mass. Hybrid inflationary
models can be an exception: the mean square of a test field can
dominate over the gauge invariant
inflaton fluctuation one on suitably chosen
parameters. We also compute the stochastic growth of quantum
fluctuations of a second field, relaxing the assumption of its zero
homogeneous value, in a generic inflationary model; as a main
result, we obtain that the equation of motion of a gauge invariant
variable associated, order by order, with a generic quantum scalar
fluctuation during inflation can be obtained only if we use the
number of e-folds as the time variable in the corresponding
Langevin and Fokker-Planck equations for the stochastic approach.
We employ this approach to derive some bounds for the case of a
model with two massive fields.
%second massive scalar field.
\end{abstract}
\maketitle

%%%%%%%%%%%%%%%%%%%%%%%%%%%%%%%%%%%%%%%%%%%%
\section{Introduction}
%%%%%%%%%%%%%%%%%%%%%%%%%%%%%%%%%%%%%%%%%%%%%%%%%%%%%%

The theory of quantum fields in an expanding universe has evolved
from its pioneering years \cite{BD} into a necessary tool in order
to describe the Universe on large scales. The de Sitter background
- characterized by the Hubble parameter $H\equiv \dot a/a$ being
constant in time (for the flat spatial slice), where $a(t)$ is the
scale factor of a Friedmann-Robertson-Walker (FRW) cosmological
model - has been the main arena to compute quantum effects
even before becoming a pillar of our understanding of the
early inflationary stage and of the recent acceleration of the
Universe.

However, while $|\dot H|\ll H^2$ for any inflationary model, $\dot
H$ may not become zero in a viable model, apart from some isolated
moments of time. Indeed, the standard slow-roll expression for the
power spectrum of the adiabatic mode of primordial scalar
(density) perturbations becomes infinite, i.e., meaningless, if
$\dot H$ becomes zero during inflation.\footnote{This is why the
statement sometimes found in literature that, for $H=const$, a
flat (Harrison-Zeldovich) $n_s=1$ spectrum is generated is also
meaningless. Actually, it is a $V(\phi)\propto \phi^{-2}$ inflaton
potential that leads to $n_s=1$ (and $r\not= 0$) in the slow-roll
approximation, see \cite{S05} for the exact solution for $V(\phi)$
without using this approximation.} Outside the slow-roll
approximation, $\dot H$ may reach zero \cite{SYK00}, but for a
moment only. Therefore, the study of quantum effects in a nearly
de Sitter stage with $\dot H \not= 0$, in particular, when the
total change in $H$ during inflation is not small compared with
its value during the last e-folds of inflation
\cite{FMVV_I,FMVV_II,FMVV_IV,FMSVV} (see also the recent papers
\cite{UM,JMPW,JP}), is not of just pure theoretical interest.
Among the main results of previous investigations, it has been
shown that the infrared growth of minimally coupled scalar fields
with a zero or small mass $m\ll H$ in a background with a
practically constant $H$ \cite{L82,S82,VF82} occurs for massive
fields when $H$ changes significantly during inflation
\cite{FMVV_I,FMVV_II}, and that the stochastic approach,
originally mainly applied to a new inflationary type background
with a small change in $H$ during inflation \cite{S86}, also works
in realistic chaotic type inflationary space-times \cite{FMSVV}
(here we do not discuss its application to eternal inflation
\cite{L86,LLM94} and to interactions in a de Sitter background
\cite{TW,SY,KO}). A particle production in a realistic
inflationary background is so different from the corresponding one
in the de Sitter space-time that it has prompted us to reconsider
the amplification of nearly massless minimally coupled scalar
fields in inflation with a quadratic potential \cite{FMVV_I,FMSVV}
and compare it with the dynamics driven by a scalar field
condensate.

In this paper we wish to tackle in more detail the moduli problem
issue. On one hand we wish to extend our results in \cite{FMSVV}
to different inflationary models and to different types of test
fields, not only to massive minimally coupled scalar fields, but
also to massless non-minimally coupled scalar fields and moduli
with an effective mass $\propto H^2$. We investigate effects
${\cal O} (\dot H)$ and we therefore need to consider both these
extensions. On the other hand, it is known that the mean square of
gauge invariant inflaton fluctuations grows \cite{FMVV_II} and the
stochastic description for this effect for a general potential has
been established \cite{FMSVV}; it is therefore interesting to
compare the quantum amplification of test fields not only with the
background inflaton dynamics, but also with the stochastic growth
of gauge invariant inflaton fluctuations. This comparison aims for
a self-consistent understanding of quantum foam during inflation.

We then discuss the diffusion equation for general scalar fluctuations
in a generic model of inflation.
On using the results obtained by field theory methods,
we show that the diffusion equation for the gauge invariant variable
associated with this generic scalar fluctuation should be
formulated in terms of the number of e-folds $N$.

The paper is organized as follows.
In Section II we choose four representative cases of
the inflationary {\em ``zoo''} on which we focus in this paper.
In Section III we compute the
stochastic growth of a minimally coupled scalar with a small
mass, a modulus with an effective mass $ \propto H^2$, and
a massless non-minimally coupled scalar in the test field
approximation for the four inflationary models considered.
In Section IV we review the result obtained in Ref. \cite{FMSVV}
for gauge invariant inflaton fluctuations and compare it with
the growth of the test fields in the four different inflationary models
considered.
In Section V we examine the stochastic approach for a two field model
for two generic self-interacting potentials, and
we compare our solution, obtaining some constraints on the
parameters, for a particular two quadratic field model.
In Section VI we illustrate our conclusions.

Our paper does not include  a derivation of  the stochastic approach. Indeed
an introduction of the stochastic method is given in Refs. [14,18] and
a comparison between stochastic methods and quantum field theory results 
is done in Refs. [14,18] and in our previous paper [7].

%%%%%%%%%%%%%%%%%%%%%%%%%%%%%%%%%%%%%%%%%%%%
\section{Inflationary Models}
%%%%%%%%%%%%%%%%%%%%%%%%%%%%%%%%%%%%%%%%%%%%

The detailed evolution of the expansion during the accelerated stage
depends on the inflaton potential and so does the growth of quantum fluctuations.
For this reason we consider in the following
four different potentials which are representative of the {\em ``inflationary zoo''}.
Since we shall study the growth of quantum fluctuations as a function of the number of e-folds
\be
N = \log \frac{a(t)}{a(t_i)}
\ee
we shall give the evolution of the Hubble parameter as a function of $N$.

The first obvious model is chaotic quadratic inflation, which we have also
used in our previous investigations \cite{FMVV_I,FMVV_II,FMVV_IV}:
\be
V(\phi) = \frac{m^2}{2} \phi^2 \,.
\ee
During the slow-roll trajectory
we have:
\begin{eqnarray}
H^2 &\simeq& H_i^2 - \frac{2}{3} m^2 N \label{htraj_quadratic} \\
\dot \phi &\simeq& - \sqrt{\frac{2}{3}} m M_{\rm pl} \label{phitraj_quadratic}\,,
\end{eqnarray}
where $M_{pl}^{-2}=8\pi G$ (these formulas were obtained already
in \cite{S78} in the context of a closed bouncing FRW universe
with two quasi-de~Sitter stages during contraction and expansion).
Let us note that in the numerical results presented in the figures
all dimensional quantities have been rescaled w.r.t.
$m_{pl}=M_{pl} \sqrt{8\pi}$. We then consider the case of a
quadratic potential (of arbitrary sign) uplifted with an offset
$V_0$: 
\be 
V(\phi) = V_0 \pm \frac{M^2}{2} \phi^2 \,.
\label{hybridorsmall} 
\ee 
With the positive sign the potential in
Eq. (\ref{hybridorsmall}) is an approximation for the simplest
model of hybrid inflation well above the scale of the end of
inflation; in this case $\phi$ decreases during the inflationary
expansion. With the negative sign the potential in Eq.
(\ref{hybridorsmall}) is a simple small field inflation model,
again far from the end of inflation; in this case $\phi$ increases
during the inflationary expansion. In the following we use the
approximate solution for the square of the inflaton as: 
\be 
\phi^2(N) \simeq \phi_i^2 e^{ \mp 2 \frac{M^2 M_{\rm pl}^2}{V_0}
(N-N_i)} \,, 
\label{phiapprox} 
\ee 
valid when $H \simeq \sqrt{V_0/(3 M_{pl}^2)}$, i.e.
$\frac{M^2}{V_0}|\phi^2-\phi_i^2|\ll 1$~\footnote{For a double
well SSB
  potential
$V=\lambda(\phi^2-\bar{\phi}^2)^2$, which may be approximated by
  Eq. (\ref{hybridorsmall}) for small field values, the condition becomes
  $|(\phi^2-\phi_i^2)/\bar{\phi}^2| \ll 1$ together with
  $\phi_i^2/\bar{\phi}^2\ll 1$, which again means that $V(\phi)$ is always considered
far from the minima.}. The exact expression can be given in the implicit form
$N-N_i=\mp \frac{V_0}{2 M^2  M_{\rm pl}^2} \log(\frac{\phi^2}{\phi_i^2})-
\frac{\phi^2-\phi_i^2}{4 M_{\rm pl}^2}$.

As another large field inflationary model we consider an exponential potential
\be
V = V_0 e^{-\frac{\lambda}{M_{\rm pl}} \phi} \,.
\label{exponential}
\ee
This potential leads to a power-law expansion given by:
\be
a(t)=\left(\frac{t}{t_i}\right)^p\,\,\,\,\,\,,\,\,\,\,\,\,H(t)=\frac{p}{t}
\ee
with $p = 2/\lambda^2$ \cite{exponential} (we consider $a (t_i) =1$ for the sake
of simplicity).
Such a solution is stable for $p>1$ and the
slow-roll conditions are well satisfied for $p >> 1$.
In particular one obtains
\begin{eqnarray}
H &=& H_i \, e^{-N/p} = \frac{p}{t_i} \, e^{-N/p} \label{htraj_exp} \\
\phi &=& \sqrt{\frac{2}{p}} M_{pl} N + \phi_i \,.
\end{eqnarray}

%%%%%%%%%%%%%%%%%%%%%%%%%%%%%%%%%%%%%%%%%%%%%%%%%
\section{Growth of test fields with small effective mass
in the stochastic approach}
%%%%%%%%%%%%%%%%%%%%%%%%%%%%%%%%%%%%%%%%%%%%%%%%%

We shall consider a test scalar field with a small effective mass and a zero homogeneous
expectation value on an inflationary background driven by an inflaton with
potential $V(\phi)$ in the slow-roll approximation. 
The evolution equation for the renormalized mean square 
$\langle \chi^2 \rangle_{\rm REN}$ (the pedix ${\rm REN}$ will denote 
renormalized in the following)
in the next three subsections
(Eqs. 11, 16 and 22) follows in a straightforward manner from our previous paper [7].
Note that the right hand side of Eqs. (11,16,22) representing the contribution of created
fluctuations ("particles") is obtained under the natural
assumption of the absence of particles in the in-vacuum state,
more exactly that each Fourier mode ${\bf k}$ of the quantum field
$\chi$ was in the adiabatic vacuum state deep inside the Hubble
radius and long before the first Hubble radius crossing during
inflation, i.e., when its energy $\omega=k/a(t)$ was much larger than
$H (t)$. \footnote{This assumption also means the absence of the
so called trans-Planckian particle creation, see \cite{KST07} and
references therein for discussion why trans-Planckian particle creation
should be absent in the standard QFT.} So, the explicit time-asymmetry
of Eq. (11) shows that this in-vacuum is not de Sitter invariant; it
is unstable and creation of fluctuations (particles) of light scalar
fields, as well as metric perturbations, takes place. In turn, the
cause of this instability may be finally traced to the expansion of
the Universe.

%%%%%%%%%%%%%%%%%%%%%%%%%%%%%%%%%%%%%%%%%%%%%%%%%
\subsection{Growth of scalar fields with $m_\chi^2$
in the stochastic approach}
%%%%%%%%%%%%%%%%%%%%%%%%%%%%%%%%%%%%%%%%%%%%%%%%%%%%%%%%%%%%%%%%%%%

The stochastic equation is:
\be
\frac{d \langle \chi^2 \rangle_{\rm REN}}{d N}
+ \frac{2 m^2_\chi}{3 H^2 (N)} \langle \chi^2 \rangle_{\rm REN} =
\frac{H^2 (N)}{4 \pi^2} \,.
\label{stochastic_mass}
\ee
Its general solution is
\ba
\langle \chi^2 \rangle_{\rm REN} =\!\!\!\!\!\!\!
&& \left( \langle \chi^2 \rangle_{\rm REN} (N_i)
+ \int^N \!\!\!dn \frac{H^2(n)}{4 \pi^2} e^{\int^n \frac{2 m^2_\chi}{3 H^2 ({\tilde n})} d{\tilde n}}
\right)\times \nonumber\\
&& e^{-\int^N \frac{2 m^2_\chi}{3 H^2 (n)} dn} \,,
\label{solution_mass}
\ea
which is just the integral form of Eq. (13) of Ref. \cite{FMSVV}
generalized to an arbitrary inflaton potential.

For the quadratic inflaton case we report here the solution given in
Ref. \cite{FMSVV}~\footnote{For the particular value $m_\chi^2 = 2 m^2$
we obtain
\be
\langle \chi^2 \rangle_{\rm REN} = \frac{3 H^4}{4
\pi^2 m^2} \log \left(\frac{H_i}{H}\right) \nonumber\,.
\ee}:

\begin{eqnarray}
\!\!\!\!\!\!\!\!\!\!\!\langle \chi^2 \rangle_{\rm REN}
&=& \!\!\frac{3 H^{2 \frac{m^2_\chi}{m^2}}}{8
\pi^2 (2 m^2 - m_\chi^2)} ( H_i^{4-2 \frac{m^2_\chi}{m^2}} \!
- \!H^{4-2 \frac{m^2_\chi}{m^2}} ) \,,
\label{soldiff}
\end{eqnarray}
where we have assumed $\langle \chi^2 \rangle_{REN} (N_i) = 0$ (we shall
adopt the same choice afterwards if not otherwise stated).
We then consider the potentials in Eq. (\ref{hybridorsmall}), in the
lowest
order
approximation; that is, for $V \simeq V_0=3 H_0^2 M_{pl}^2$, we have
\be
\langle \chi^2 \rangle_{\rm REN} \simeq  \frac{3 H_0^4}{8 \pi^2 m_\chi^2}
\Bigl(1-e^{-\frac{2 m_\chi^2}{3 H_0^2}
N}\Bigr) \,.
\label{massivemod}
\ee

Let us note that  the  corrections induced by a non-zero
$M^2\phi^2 /V_0$ term are typically small both for the case of
hybrid inflation as well as for small field inflation (as long as
the field does not grow too much due to instability). The
corresponding analytic expressions, obtained using
Eq.~\eqref{phiapprox}, can be written in terms of hypergeometric
functions but we do not report them here. For the exponential
potential we obtain:
%\begin{widetext}
\begin{eqnarray}
\langle \chi^2 \rangle_{\rm REN} &=& \frac{p}{8 \pi^2} H_i^2
\exp\left(-\frac{p}{3}\frac{m^2_\chi}{H^2}\right)
\left[-\exp\left(\frac{p}{3}\frac{m^2_\chi}{H^2}\right)\frac{H^2}{H_i^2}
\right. \nonumber \\
& & \left.
+\frac{p}{3}\frac{m^2_\chi}{H_i^2}
Ei \left(\frac{p}{3}\frac{m^2_\chi}{H^2}\right)
%\right. \nonumber \\ & & \left.
+\exp\left(\frac{p}{3}\frac{m^2_\chi}{H_i^2}\right)
\right. \nonumber \\
& & \left. -
\frac{p}{3}\frac{m^2_\chi}{H_i^2}
Ei \left(\frac{p}{3}\frac{m^2_\chi}{H_i^2}\right)\right] \,,
\end{eqnarray}
%\end{widetext}
where $E_i$ is the exponential integral function (see, for example, \cite{GR}).

%%%%%%%%%%%%%%%%%%%%%%%%%%%%%%%%%%%%%%%%%%%%%%%%%
\subsection{Growth of moduli fields with $m_\chi^2 = c H^2$
in the stochastic approach}
%%%%%%%%%%%%%%%%%%%%%%%%%%%%%%%%%%%%%%%%%%%%%%%%%%%%%%%%%%%%%%%%%%%
If $|c|\ll 1$, the stochastic equation takes the form: \be \frac{d
\langle \chi^2 \rangle_{\rm REN}}{d N} + \frac{2 c}{3} 
\langle \chi^2 \rangle_{\rm REN} = \frac{H^2 (N)}{4 \pi^2} \,.
\label{stochastic_moduli} \ee Its general solution is \be \langle
\chi^2 \rangle_{\rm REN} = \left( \langle \chi^2 \rangle_{\rm REN}
(N_i) + \int^N d n \frac{H^2(n)}{4 \pi^2} e^{\frac{2}{3} c n}
\right) e^{-\frac{2}{3} c N} \,.
\label{solution_moduli} \ee

For $V(\phi)= m^2 \phi^2 /2$ we obtain: \be \langle \chi^2
\rangle_{\rm REN} = \frac{m^2}{6\pi^2}\Bigl[
\Bigl(1-e^{-\frac{2}{3} c N}\Bigr)\left(\frac{9}{4c^2}+
\frac{3}{2c}N_T\right)-\frac{3}{2c}N\Bigr]\,, \ee where
$N_T=\frac{3H_i^2}{2m^2}$ is equal to maximal number of possible
e-folds in this chaotic model. In the limiting case $c\to 0$ and
at the end of inflation ($N=N_T-3/2$), we recover the
result\footnote{This corresponds to the massless
  limit of moduli production computed in  Eq. (15) of \cite{FMSVV} for $\alpha \to 0$.}:
\be
\langle \chi^2 \rangle_{\rm REN} \simeq
\frac{m^2}{12\pi^2}N_T^2=\frac{3H_i^4}{16\pi^2m^2} \,,
\label{massless_moduli}
\ee

For the potential in Eq. (\ref{hybridorsmall}), we consider the lowest
approximation as in the previous subsection.
Therefore the result can be simply obtained by substituting
$m_\chi^2 = c H_0^2$ in Eq.~\eqref{massivemod}:
\be
\langle \chi^2 \rangle_{\rm REN} = \frac{3 H_0^2}{8\pi^2 c}
\Bigl(1-e^{-\frac{2}{3} c N}\Bigr) \,.
\ee

For the power-law inflation case we obtain:
\be
\langle \chi^2 \rangle_{\rm REN} = \frac{p}{8 \pi^2} H_i^2 \left(c
\frac{p}{3}
-1\right)^{-1}\left(e^{-2 \frac{N}{p}}-e^{-\frac{2}{3} c N}\right)\,.
\ee

Let us note that $m^2=V_{\phi\phi}=\frac{6}{p}(1-\frac{1}{3 p})
H^2$ for this particular model, so the results above are also
valid for the case $m_{\chi}^2=\tilde c m^2$ with $\tilde
c=\frac{p}{6}(1-\frac{1}{3 p})^{-1} c$.

%%%%%%%%%%%%%%%%%%%%%%%%%%%%%%%%%%%%%%%%%%%%%%%%%%%%%%%%%%%%
\subsection{Growth of non-minimally coupled scalar fields in the stochastic approach}
%%%%%%%%%%%%%%%%%%%%%%%%%%%%%%%%%%%%%%%%%%%%%%%%%%%%%%%%%%%%%%%%%%%%%%%%%%%
The stochastic equation is now: \be \frac{d \langle \chi^2
\rangle_{\rm REN}}{d N} + 4 \xi (2 - \epsilon) \langle
\chi^2 \rangle_{\rm REN} = \frac{H^2 (N)}{4 \pi^2} \,,
\label{stochastic_nmc} \ee where $\xi$ is the non-minimal coupling
to the Ricci scalar $R$ and we assume that $|\xi|\ll 1$ (however,
$\xi N$ may be large). Indeed, the term in the action proportional
to $\xi \chi^2 R$ gives an effective time dependent mass for
$\chi$: $m_\chi^2=6\xi H^2 (2-\epsilon)$ where
$\epsilon=-\frac{\dot H}{H^2}$.

Its general solution is
\ba
\langle \chi^2 \rangle_{\rm REN} &=&
\Bigl[ \langle \chi^2 \rangle_{\rm REN} (N_i)
+ \int^N d n \frac{H^{2+4 \xi} (n)}{4 \pi^2 H_i^{4 \xi}} e^{8 \xi n}
\Bigr]\times \nonumber \\
& &
 \left( \frac{H_i}{H(N)} \right)^{4 \xi}  e^{- 8 \xi N}\,.
\label{solution_nmc}
\ea

If we again consider the chaotic scenario induced by a massive
inflaton field as in the previous subsections, the integral can be
easily computed in a closed form in terms of the exponential
integral function $E_\nu(z)$. Assuming $\langle \chi^2
\rangle_{\rm REN}=0$ initially, one finds
\begin{widetext}
\be
\langle \chi^2 \rangle_{\rm REN} \simeq
\frac{m^2}{6\pi^2} \frac{e^{8 \xi (N_T-N)}}{(N_T\!-\!N)^{2\xi}}
\Bigl[
(N_T\!-\!N)^{2+2\xi}E_{-1-2\xi}\bigl(8\xi(N_T\!-\!N)\bigr)-
(N_T\!-\!N_i)^{2+2\xi}E_{-1-2\xi}\bigl(8\xi(N_T\!-\!N_i)\bigr)
\Bigr]\,.
\ee
\end{widetext}
One can verify that in the limit $\xi \to 0$ at the end of
inflation and a fixed large value for $N_T$, the result of
Eq.~\eqref{massless_moduli} for a massless modulus is again
reobtained.

For the potential in Eq.~(\ref{hybridorsmall}) we have, in the
same approximation as in the two previous subsections, \be \langle
\chi^2 \rangle_{\rm REN} = \frac{H_0^2}{32 \pi^2 \xi}
\Bigl(1-e^{-8 \xi N}\Bigr) \,. \ee

As before, let us finish with the case of a power-law model of
inflation.
%For $\langle \chi^2 \rangle (t_i) = 0$,
We obtain:
\be
\langle \chi^2 \rangle_{\rm REN} = \frac{p}{8 \pi^2} H_i^2 \left(-2
\xi-1
+4 p \xi\right)^{-1}\left(e^{-2 \frac{N}{p}}-e^{\xi N\left(\frac{4}{p}-8
\right)}\right) \,.
\ee

%%%%%%%%%%%%%%%%%%%%%%%%%%%%%%%%%%%%%%%%%%%%%%%%%%%%%%%%%%%%%%%%%%%%%
\section{Comparison with the growth of inflaton fluctuations}
%%%%%%%%%%%%%%%%%%%%%%%%%%%%%%%%%%%%%%%%%%%%%%%%%%%%%%%%%%%%%%%%%%%%%

The results of the previous section should be compared with the
growth of gauge-invariant inflaton fluctuations $\delta \phi$, 
the Mukhanov variable
\cite{Mukhanov} which is used to canonically quantize the 
Einstein-Klein-Gordon Lagrangian.
The evolution equation for
$\langle \delta \phi^2 \rangle_{\rm REN}$ found in
\cite{FMSVV} can be re-written as:

\be
\frac{d \langle \delta \phi^2 \rangle_{\rm REN}}{d N}
+ 2 \left( \eta - 2 \epsilon \right)
\langle \delta \phi^2 \rangle_{\rm REN} =
\frac{H^2 (t)}{4 \pi^2} \,,
\label{inflaton}
\ee
where
\ba
\epsilon &=& \frac{M_{\rm pl}^2}{2}
\left( \frac{V_\phi}{V}\right)^2 \, , \nonumber \\
\eta &=& M_{\rm pl}^2 \frac{V_{\phi \phi}}{V} \,.
\ea
In Eq. (\ref{inflaton}) the positivity of $\eta - 2 \epsilon$ is
not determined by the convexity of the potential,
 i.e. $V_{\phi \phi} > 0$, as we would expect in the absence of
metric perturbations. The threshold corresponds to the following
condition on the potential: \be \frac{d}{d \phi} \left(
\frac{V_\phi}{V} \right) > 0\,. \label{potential_condition} \ee

With the use of the slow-roll expressions for the scalar spectral
index $n_s$ and the tensor-to-scalar ratio $r$, one obtains: 
\ba
n_s -1 &=& -6 \epsilon + 2 \eta \\
r &=& 16 \epsilon \,, \ea and Eq. (\ref{inflaton}) can be
rewritten as: \be \frac{d \langle \delta \phi^2 \rangle_{\rm
REN}}{d N} + \left( n_s - 1 + \frac{r}{8} \right) \langle \delta
\phi^2 \rangle_{\rm REN} = \frac{H^2 (t)}{4 \pi^2} \,.
\label{inflaton2} 
\ee 
Both Eqs. (\ref{potential_condition}) and
(\ref{inflaton2}) tell us that power-law inflation,  for which
$n_s - 1 = -r/8$ holds, lies at the threshold between two opposite
behaviors. Power-law inflation with $78 < p < 246$ is allowed at
the 95 $\%$ confidence level \cite{FHLL}. We note that Eq.
(\ref{inflaton2}) is the same for a modulus with the mass $m^2_\chi
= c H^2$ and $c = 3(n_s-1+r/8)/2$: below the power-law inflation
line inflaton fluctuations behave as a modulus with negative $c$.

\begin{figure}[h]
\centering
\includegraphics[height=.25\textwidth]{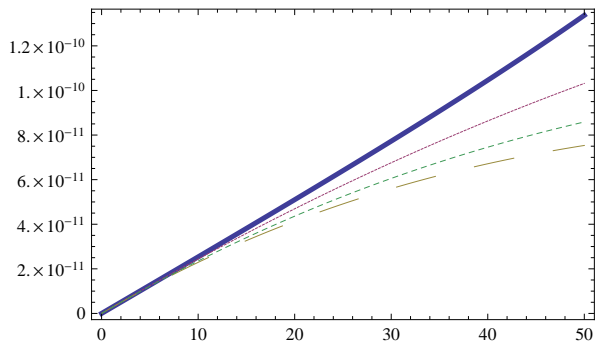}
\caption{\label{fig:quadratic}
Evolution of the mean square quantum fluctuations (in units of $m_{\rm pl}^2$)
versus the number of e-folds $N$ for the quadratic chaotic model.
For the inflationary background we have chosen the inflationary trajectory in Eq. (\ref{htraj_quadratic})
with $m=10^{-6} \, m_{\rm pl}$ and $H_i = 10 \, m$. The mean square
 gauge invariant inflaton fluctuation (thick line) dominates over those of
test fields ($m_\chi \simeq 0.3 m$ is the solid line, $c=0.02$ is the dashed line, $\xi=0.001$ is the dotted line).
}
\end{figure}

\begin{figure}[h]
\centering
\includegraphics[height=.25\textwidth]{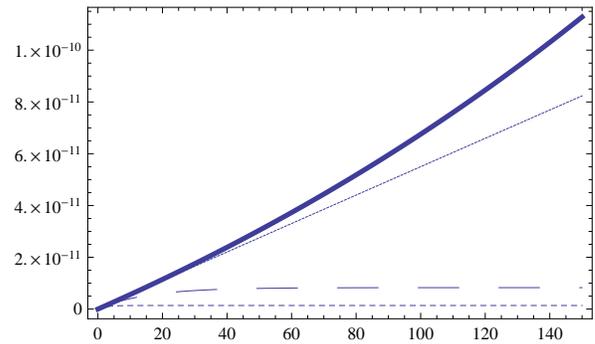}

\caption{\label{fig:small} Evolution of the mean square quantum
fluctuations (in units of $m_{\rm pl}^2$) versus the number of
e-folds $N$ for the small field inflationary model in Eq.
(\ref{hybridorsmall}). For the inflationary background we have
chosen $V_0=2.6 \times 10^{-12} m_{\rm pl}^4$, $M = 0.85 \times
10^{-6} m_{\rm pl}$ and $\phi_i = 0.3 \, m_{\rm pl}$ as
parameters. The mean square gauge invariant inflaton fluctuation
(thick line) dominates over those of test fields ($m_\chi =
10^{-2} H_0$ is the solid line, $c=0.1$ is the dashed line,
$\xi=0.05$ is the dotted line). }
\end{figure}

The solution of Eq. (\ref{inflaton}) is:
\be
\langle \delta \phi^2 \rangle_{\rm REN}
= \frac{\epsilon(N)}{4 \pi^2} \int^N d n
\frac{H^2 (n)}{\epsilon (n)} \,.
\ee
For the quadratic chaotic potential the solution was found in
\cite{FMSVV}:
\be
\langle \delta \phi^2 \rangle_{\rm REN}
= \frac{H_i^{6} - H^{6}}{8 \pi^2 m^2 H^2} \,.
\ee

For the potential in Eq. (\ref{hybridorsmall}) we obtain the following 
in the
lowest non-trivial approximation for small $M^2$-dependent
corrections in the potential: \be \langle \delta \phi^2
\rangle_{\rm REN} \simeq \pm \frac{V_0^2}{24 \pi^2 M^2 M_{\rm
pl}^4} \left[ 1 - e^{\mp \frac{2 M^2 M_{\rm pl}^2}{V_0} (N-N_i)}
\right] \,, \ee where, as discussed previously, the case with a
minus sign in the exponent refers to the hybrid model whereas the
other case is associated with the small field inflationary model.
Let us note that for a small value of the exponent (an almost
constant potential) or for very small ($N-N_i$), by expanding up
to the linear order, one obtains almost the case of a de Sitter
background ($M^2=0$), with $\langle \delta \phi^2 \rangle_{\rm
REN} $ linearly growing in $N$. In this approximation we see that
the hybrid model is characterized by $\langle \delta \phi^2
\rangle_{\rm REN} $ bounded as $N \to \infty$. This statement
remains true even after dropping our approximations (see below).

This leads, by invoking the consistency of the perturbative
expansion in field fluctuations through the condition $\langle
\delta \phi^2 \rangle_{\rm REN}\ll \phi_i^2$, to the following
hierarchy which the inflationary model has to satisfy: \be
\frac{V_0}{24 \pi^2 M_{\rm pl}^4} \ll \frac{M^2 \phi_i^2}{V_0} \ll
1 \,, \ee

We have also computed the expression for the fluctuations by
solving Eq.~\eqref{inflaton} using the expression in
Eq.~\eqref{phiapprox} with no further approximations. In this more
general case we obtain
\begin{widetext}
\be
\langle \delta \phi^2 \rangle_{\rm REN} \simeq
\pm \frac{4 V_0^2(1-y)+3M^4 \phi_i^2 y \left(4M_{pl}^2 (N-N_i) +
\phi_i^2(1-y)\right) \pm y(1-y^2)\frac{M^6}{4V_0}}
{96 \pi^2 M^2 M_{\rm pl}^4 \left(1\pm y \frac{M^2 \phi_i^2}{2V_0}\right)^2} \,,
\ee
\end{widetext}
where we have set $y=y(N)=e^{\mp\frac{2M^2 M_{\rm
pl}^2}{V_0}(N-N_i)}$. From this expression, when analyzing the
hybrid inflation case, one can notice that the fluctuations have a
maximum for a certain amount of e-folds $N$ and then decay to the
asymptotic value for a large number of e-folds. Nevertheless, such
a maximum is typically a few percent above the asymptotic value
which has been already obtained above using a more crude
approximation.

\begin{figure}[h!]
\centering
\includegraphics[height=.25\textwidth]{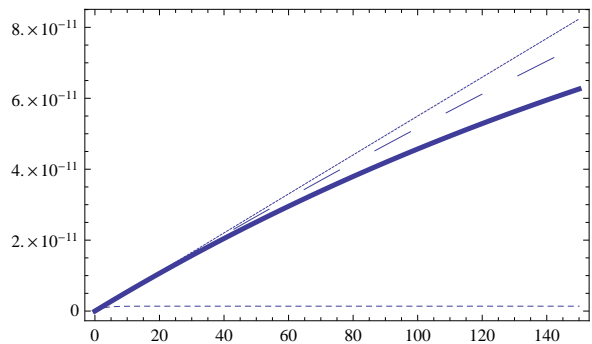}
\caption{\label{fig:hybrid} Evolution of the mean square quantum
fluctuations (in units of $m_{\rm pl}^2$) versus the number of
e-folds $N$ for the hybrid model in Eq. (\ref{hybridorsmall}). For
the inflationary background we have chosen $V_0=2.6 \times
10^{-12} m_{\rm pl}^4$, $M = 1.8 \times 10^{-6} m_{\rm pl}$ and
$\phi_i = 0.3 \, m_{\rm pl}$ as parameters. In this case the mean
square of moduli can dominate over the mean square of gauge
invariant inflaton fluctuation (thick line): the parameters chosen
are $m_\chi = 10^{-2} H_0$ (solid line), $c=0.002$ (dashed line),
$\xi=0.05$ (dotted line). }
\end{figure}

For the exponential potential we find
 \be
\langle \delta \phi^2 \rangle_{\rm REN} = \frac{p}{8 \pi^2}(H_i^2-H^2) \,,
\ee

which at late times (see also \cite{Marozzi}) becomes:
\be
\langle \delta \phi^2 \rangle_{\rm REN} = \frac{p H_i^2}{8 \pi^2} \,.
\ee

\begin{figure}[h!]
\centering
\includegraphics[height=.25\textwidth]{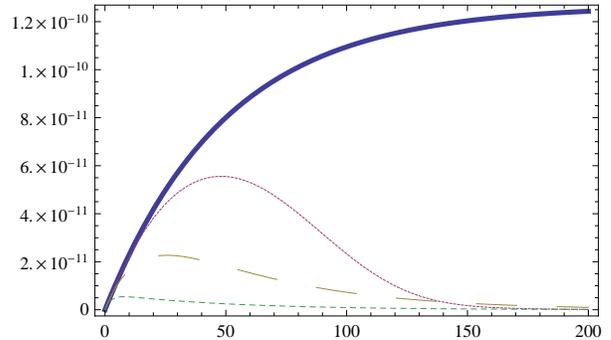}
\caption{\label{fig:powerlaw}
Evolution of the mean square quantum fluctuations (in units of $m_{\rm pl}^2$)
versus the number of e-folds $N$ for the exponential potential.
For the inflationary background we have chosen the inflationary trajectory in Eq. (\ref{htraj_exp})
with $p=100$ and $t_i = 10^7 \, m_{\rm pl}^{-1}$. The mean square gauge
 invariant inflaton fluctuation (thick line) dominates over those of
test fields ($m_\chi = 10^{-6} \, m_{\rm pl}$ is the solid line, $c=0.1$ is
the dashed line, $\xi=0.05$ is the dotted line).
}
\end{figure}

We show in Figures \ref{fig:quadratic}, \ref{fig:small},
\ref{fig:hybrid} and \ref{fig:powerlaw}, one for each inflaton
potential investigated, the mean square of quantum fluctuations of
the three types of test fields together with the gauge invariant
inflaton ones. For the hybrid model, quantum fluctuations of test
fields with a small effective mass can dominate the
gauge-invariant inflaton one, because of the presence of the
leading constant term in the potential.

%%%%%%%%%%%%%%%%%%%%%%%%%%%%%%%%%%%%%%%%%%%%%%%%%%%%%%%%%%%%%%%%%%%%%%%%%%%%
\section{Growth of quantum fluctuation in two field inflationary models}
%%%%%%%%%%%%%%%%%%%%%%%%%%%%%%%%%%%%%%%%%%%%%%%%%%%%%%%%%%%%%%%%%%%%%%%%%%%%

We now wish to consider a two field model in which an inflaton
$\phi$ and a minimally coupled scalar field $\chi$ are present
(see \cite{LMY} for a different approach to the moduli problem).
We shall neglect the $\chi$ energy density and pressure in the
background FRW equations. We expand to second order in the uniform curvature 
gauge (UCG), in
which the inflaton fluctuation $\varphi$ coincides with the
gauge-invariant Mukhanov variable, the Einstein and Klein-Gordon
equations.

In the test field expansion $\chi(\vec{x},
t)=\chi_0(t)+\chi^{(1)}(\vec{x}, t) +\chi^{(2)}(\vec{x}, t)+...$,
the homogeneous term satisfies

\be
\ddot{\chi}_0+3 H
\dot{\chi}_0+\bar{V}_\chi=0 \,,
\label{QFTorder0chi}
\ee

while fluctuations satisfy, order by order, for the leading order
in the slow-roll approximation and in the long-wavelength limit
(neglecting vector and tensor contributions), the following
equations:

\be
3 H \dot{\chi}^{(1)}
+\bar{V}_{\chi\chi}\chi^{(1)}=2 \frac{H_\phi}{H} \bar{V}_\chi \varphi^{(1)} \,,
\label{QFTorder1chi}
\ee

\ba
& & 3 H \dot{\chi}^{(2)}
+\bar{V}_{\chi\chi}\chi^{(2)}
\nonumber\\
& &
= 2\frac{H_\phi}{H} \bar{V}_\chi \varphi^{(2)}
+\left[\frac{H_{\phi\phi}}{H}
-3 \left(\frac{H_\phi}{H}\right)^2\right]
\bar{V}_\chi \varphi^{(1) 2}
 \nonumber\\
& & +2 \bar{V}_{\chi\chi}  \frac{H_\phi}{H} \varphi^{(1)}
\chi^{(1)}
-\frac{\bar{V}_{\chi\chi\chi}}{2} \chi^{(1) 2} \,.
\label{QFTorder2chi}
\ea

Following the consideration in section VI of Ref. \cite{FMSVV}, we
wish to investigate which time variable in the stochastic equation
should be chosen to re-obtain, order by order, the equation of
motion for the test field $\chi$ starting from \be \frac{d \chi}{d
t}=-\frac{\bar{V}_\chi}{3 H(\phi)} \ee and expanding order by
order. For a general time variable $\tau=\int H(t)^n dt$, the
equation becomes \be \frac{1}{H(t)^n}\frac{d \chi}{d
t}=-\frac{\bar{V}_\chi}{3 H(\phi)^{n+1}} \label{StartingEqchi}\,.
\ee As before, expanding to leading order in the slow-roll
approximation, we obtain the following equation to the first and
second order: \be \frac{d \chi^{(1)}}{d t}=-\frac{1}{3 H}
\bar{V}_{\chi\chi} \chi^{(1)} +\frac{1}{3}(n+1) \frac{H_\phi}{H^2}
\bar{V}_{\chi} \varphi^{(1)} \,, \ee
\begin{widetext}
\be
\frac{d \chi^{(2)}}{d t}\!=\!-\frac{1}{3 H} \bar{V}_{\chi\chi} \chi^{(2)}
+\frac{1}{3}(n+1) \frac{H_\phi}{H^2} \bar{V}_{\chi} \varphi^{(2)}
-\frac{1}{6 H} \bar{V}_{\chi\chi\chi} \chi^{(1) 2}+\frac{1}{3}(n+1)
\frac{H_\phi}{H^2} \bar{V}_{\chi\chi} \varphi^{(1)} \chi^{(1)}-\frac{1}{6}
(n+1) \bar{V}_{\chi}\left[-\frac{H_{\phi\phi}}{H^2}+(n+2)
\frac{H_{\phi}^2}{H^3}\right] \varphi^{(1) 2} \,.
\ee
\end{widetext}
As is easy to verify, we recover the former result only for $n=1$.
So for the case of a test scalar field, evolving in a FRW inflaton
driven space-time, in the UCG the right time to consider is the
number of e-folds $N=\int H(t) dt$, this recovers the result
obtained in \cite{FMSVV} for the inflaton fluctuations. As for the
case of the standard Mukhanov variable $Q$, which is defined,
order by order, as the value of the inflaton perturbation in the
UCG, we can define a generic gauge-invariant Mukhanov variable
$Q_\chi$, associated with the perturbation of $\chi$, as the value
that this perturbation has in the UCG. 
In this way $Q_\chi^{(n)}=\chi^{(n)}$ in the UCG and, as for the variable $Q$
in \cite{FMSVV}, the equations above can be regarded as the
gauge-invariant equations of motion, to first and second
order, of this new Mukhanov variable, where one replaces
$\chi^{(n)}$ with $Q_\chi^{(n)}$ and $\varphi^{(n)}$ with
$Q^{(n)}$.

As for the case of the Mukhanov variable, this result can be
considered as a starting point to study the fluctuations of $\chi$
in the stochastic approach for an arbitrary potential $\bar{V}$ in
the described background. The correct stochastic differential
equation is obtained with respect to the number of e-folds ($d N =
H(t) d t$) which appears to be the right evolution parameter. One
starts from the slow-roll approximation to the Heisenberg
equation, which can be interpreted in a general non-perturbative
sense, for the large-scale quantum field $\chi$ \ba \frac{d}{d N}
\chi &=& -\frac{1}{3H^2} \bar{V}_\chi + \frac{1}{H} f_\chi\,, \nonumber \\
%\langle f_\chi(N_1) f_\chi(N_2) \rangle &=& \frac{H^4}{4\pi^2}
%\delta(N_1 -N_2) \, . 
\langle f_\chi(N_1 \,, {\bf x_1}) f_\chi(N_2 \,, {\bf x_2}) \rangle &=& \frac{H^4}{4\pi^2}
\delta(N_1 -N_2) 
\frac{\sin (|{\bf x_1} - {\bf x_2}|)}{|{\bf x_1} - {\bf x_2}|} \,, \nonumber
%\label{chi_general_stoch} 
\ea 
where $f_\chi$
is the stochastic noise term given, to the leading order in the
slow-roll approximation, by \ba f_\chi(t, {\bf x})&=&\epsilon a
H^2 \int \frac{d^3 k}{(2 \pi)^{3/2}} \delta (k-\epsilon a H) \left
[ \hat{b}_k \chi_k(t) e^{- i {\bf k \cdot x}} \right.
\nonumber \\
& & \left. + \hat{b}^\dagger_k \chi^*_k(t) e^{+i {\bf k \cdot x}}
\right]\,. \ea 
Thus, on expanding to the second order, one obtains
the following stochastic equations for $\chi^{(1)}$ and
$\chi^{(2)}: $ 
\be 
\frac{d \chi^{(1)}}{d t}=-\frac{1}{3 H}
\bar{V}_{\chi\chi} \chi^{(1)} +\frac{2}{3} \frac{H_\phi}{H^2}
\bar{V}_{\chi} \varphi^{(1)}+f_\chi \,, \label{Stochchi1} \ee
\begin{widetext}
\be
\frac{d \chi^{(2)}}{d t}=-\frac{1}{3 H} \bar{V}_{\chi\chi} \chi^{(2)}
+\frac{2}{3}\frac{H_\phi}{H^2} \bar{V}_{\chi} \varphi^{(2)}
-\frac{1}{6 H} \bar{V}_{\chi\chi\chi} \chi^{(1) 2}+\frac{2}{3}
\frac{H_\phi}{H^2} \bar{V}_{\chi\chi} \varphi^{(1)} \chi^{(1)}-\frac{1}{3}
\bar{V}_{\chi}\left[-\frac{H_{\phi\phi}}{H^2}+3
\frac{H_{\phi}^2}{H^3}\right] \varphi^{(1) 2}+\frac{H_\phi}{H} \varphi^{(1)}
f_\chi \,.
\label{Stochchi2}
\ee
\end{widetext}
Let us consider the first order stochastic equation. Its general
solution with the zero initial condition is given by \be
\chi^{(1)}=\bar{V}_\chi \int_{t_i}^{t} \left(\frac{2}{3}
\frac{H_\phi}{H^2}
\varphi^{(1)}+\frac{f_\chi}{\bar{V}_\chi}\right)d\tau \,. \ee
Taking into account that $\langle \varphi^{(1)} f_\chi \rangle=0$,
it is easy to derive the expression for the mean square of the
first order fluctuation $\langle (\delta \chi^{(1)})^2 \rangle$ :
\begin{widetext}
\begin{eqnarray}
\langle \chi^{(1) 2}\rangle &=& \bar{V}_\chi^2 \int^t_{t_i} d\tau \int^t_{t_i}
d\eta \left[\frac{4}{9} \frac{H_\phi(\tau)}{H(\tau)^2}
\frac{H_\phi(\eta)}{H(\eta)^2} \langle \varphi^{(1)}(\tau)\varphi^{(1)}(\eta)
\rangle+\frac{1}{\bar{V}_\chi(\tau)}\frac{1}{\bar{V}_\chi(\eta)}
\langle f_\chi(\tau) f_\chi(\eta) \rangle \right] \nonumber \\
 &=& \frac{\bar{V}_\chi^2}{4 \pi^2} \int^t_{t_i} d\tau \left[
\frac{H(\tau)^3}{\bar{V}_\chi(\tau)^2}-\frac{4}{9 M_{pl}^2} \int^\tau_{t_i}
\!d\eta \, \frac{\dot{H}(\tau)}{H(\tau)^3}\frac{\dot{H}(\eta)}{H(\eta)^3}
\int^\eta_{t_i} \!d\sigma \,\frac{H(\sigma)^5}{\dot{H}(\sigma)}\right] \,,
\label{Corr_chi}
\end{eqnarray}
where we have used the stochastic solution \be
\varphi^{(1)}=\frac{V_\phi}{V}\int_{t_i}^t d\tau
\left(\frac{V}{V_\phi}f_\phi\right) \,, \ee with $f_\phi$ being
the stochastic noise for the inflaton defined analogously to
$f_\chi$. Similarly, one can obtain the following solution for a
vacuum expectation value of the second order fluctuation: \be
\langle \chi^{(2)}\rangle = \bar{V}_\chi \int^t_{t_i} d\tau\left[
\frac{2}{3}\frac{H_\phi}{H^2} \langle \varphi^{(2)}\rangle
-\frac{1}{6 H} \frac{\bar{V}_{\chi\chi\chi}}{\bar{V}_\chi} \langle
\chi^{(1) 2}\rangle+\frac{2}{3} \frac{H_\phi}{H^2}
\frac{\bar{V}_{\chi\chi}}{\bar{V}_\chi} \langle \varphi^{(1)}
\chi^{(1)}\rangle+\frac{1}{3} \left(\frac{H_{\phi\phi}}{H^2}-3
\frac{H_{\phi}^2}{H^3}\right) \langle \varphi^{(1) 2}\rangle
\right] \,, \label{chi2vev} \ee where, to expand further, we
should substitute Eq.~(\ref{Corr_chi}), the result
 for $\langle \varphi^{(2)}\rangle$ and $\langle \varphi^{(1) 2}\rangle$
obtained in \cite{FMSVV} and
\be
\langle \varphi^{(1)} \chi^{(1)}\rangle=-\frac{\bar{V}_\chi}{12 \pi^2}
\frac{\dot{\phi}}{H M_{pl}^2} \int^t_{t_i} d\tau \int^t_{\tau} d\eta
\left[\frac{H(\tau)^5}{\dot{H}(\tau)} \frac{\dot{H}(\eta)}{H(\eta)^3}\right] \,.
\label{chi_phi}
\ee

\end{widetext}

%%%%%%%%%%%%%%%%%%%%%%%%%%%%%%%%%%%%%%%%%%%%%%%%%%%%%%%%%%%%%%%%%%%%%

\subsection{A working example: two field quadratic model}
%%%%%%%%%%%%%%%%%%%%%%%%%%%%%%%%%%%%%%%%%%%%%%%%%%%%%%%%%%%%%%%%%%%%%

Let us now consider the particular case
$V(\phi)=\frac{m^2\phi^2}{2}$ and
$\bar{V}(\chi)=\frac{m_\chi^2\chi^2}{2}$. Classical slow-roll
inflation in this model and the evolution of small perturbations
in it were calculated in \cite{PS92}, but here we take into
account the backreaction of the generated quantum fluctuations of
these scalar fields on the evolution of their background values.
By solving the background equations, one obtains the following
zero order solution for the test field $\chi$ \be
\chi^{(0)}(t)=\chi^{(0)}(t_i)\left(
\frac{H(t)}{H(t_i)}\right)^{\frac{m_\chi^2}{m^2}} \,. \ee It
remains a test field for the whole duration of the inflation era
if \be \chi^{(0)}(t_i)^2 \ll
\left[1+\frac{\alpha}{9}\frac{m^2}{H^2}\right]^{-1}
\frac{1}{\alpha}\left(\frac{H}{H_i}\right)^{2-2\alpha}6
\frac{H_i^2}{m^2} M_{pl}^2 \ee for any value of $H$ (where
$\alpha=\frac{m_\chi^2}{m^2}$). For the case $\alpha \ll 1$, we
obtain the following limiting condition at the end of inflation
($H \simeq m$):
\be \chi^{(0)}(t_i)^2 \ll \frac{6}{\alpha}
M_{pl}^2 \label{ModuliCond} \ee and for this particular background
we can solve Eq.(\ref{Corr_chi}) obtaining
\begin{eqnarray}
\langle \chi^{(1) 2}\rangle &=& \frac{3 H^{2 \alpha}}{8
\pi^2 m^2 (2-\alpha)} ( H_0^{4-2 \alpha} \!- \!H^{4-2 \alpha} )+ \nonumber \\
 & & \!\!\!\!\!\!\!\!\!\!\!\!\!\!\!\!
-\frac{\alpha^2}{48 \pi^2}\frac{\chi^{(0)}(t_i)^2}{M_{pl}^2}
\left(\frac{H}{H_i}\right)^{2 \alpha}\frac{1}{H^4}\left(H^2-H_i^2\right)^3\,.
\label{CorrChiPart}
\end{eqnarray}
Thus, one obtains the term already considered in \cite{FMSVV} plus
a new term which depends on the background value of $\chi$. At the
end of inflation, the leading value of this second term is
negligible with respect to the leading value of the first one, for
$\alpha<2$, if \be \chi^{(0)}(t_i)^2 \ll
\frac{18}{2-\alpha}\frac{1}{\alpha^2} \frac{M_{pl}^2 m^2}{H_i^2}
\,. \label{CondNeg} \ee 
This condition is different from, and can be
stronger than, the condition (\ref{ModuliCond}). If we consider
the particular case $\alpha\ll 1$ and require that
(\ref{ModuliCond}) implies (\ref{CondNeg}), we obtain the
following condition on $\alpha$: \be \alpha \ll \frac{3}{2}
\frac{m^2}{H_i^2}\,. \ee Analogously we can evaluate
Eq.(\ref{chi2vev})
\begin{eqnarray}
\langle \chi^{(2)} \rangle &=&\frac{\alpha}{8 \pi^2}
\frac{\chi^{(0)}(t_i)}{M_{pl}^2} \left(\frac{H}{H_i}\right)^{\alpha}
\left[-\frac{H_i^6}{H^4}\frac{1-\alpha/2}{6}+\frac{H_i^4}{H^2}
\frac{1-\alpha}{4} \right. \nonumber \\
& & \left.
+H_i^2 \frac{\alpha}{4}
-H^2\frac{1+\alpha}{12}\right] \,,
\end{eqnarray}
which to leading order gives
\be
\langle \chi^{(2)} \rangle =-\frac{\alpha}{48 \pi^2}
\frac{\chi^{(0)}(t_i)}{M_{pl}^2} \left(\frac{H_i}{H}\right)^{4-\alpha} H_i^2
\left(1-\frac{\alpha}{2}\right) \,.
\ee

%%%%%%%%%%%%%%%%%%%%%%%%%%%%%%%%%%%%%%%%%%%%%%%%%%%%%%%%%%%%%%%%%%%%%%
\section{Conclusions}
%%%%%%%%%%%%%%%%%%%%%%%%%%%%%%%%%%%%%%%%%%%%%%%%%%%%%%%%%%%%%%%%%%%%%%

Motivated by previously found differences for gravitational
particle production in the de Sitter background and in realistic
inflationary models with $\dot H \ne 0$, we have studied in detail
the growth of quantum fluctuations for the latter case. We have
selected four different potentials as representative examples of
the {\em inflationary zoo} and different types of nearly massless
fluctuations, including inflaton ones. We have rewritten in Eq.
(\ref{inflaton2}) the diffusion equation for the gauge invariant
inflaton fluctuations found in \cite{FMSVV}, emphasizing the role
of the slope of the spectrum of curvature perturbations $n_s$ and
of the tensor-to-scalar ratio $r$, i.e. the relevant observable
quantities.

We have found that for most of the inflationary models, the mean
square of the gauge invariant inflaton fluctuations dominates over
those of moduli with a non-negative effective mass. Hybrid
inflationary models can be an exception: the mean square of a test
field can dominate over that of the gauge invariant inflaton
fluctuations on choosing parameters appropriately. Our findings
show that the understanding of inflaton dynamics including its
quantum fluctuations is more important than the moduli problem in
most of the inflationary models.

We have then discussed the stochastic approach for general scalar
fluctuations, which may have a non zero homogeneous mode, in a
generic model of inflation. We show, by using the field theory
results as a guideline, that the stochastic equations for the
gauge invariant variable associated with such scalar fluctuations
are naturally formulated as a flow in terms of the number of
e-folds $N$. Finally we have studied the particular case of a
massive inflaton and a second massive scalar field $\chi$, for
which we show how to extract some bounds for the homogeneous mode
of $\chi$ and for its mass.

\vspace{1cm}

%%%%%%%%%%%%%%%%%%%%%%%%%%%%%%
{\bf Acknowledgements}
%%%%%%%%%%%%%%%%%%%%%%%%%%%%%%%%

G.M. was supported by the GIS "Physique des Deux Infinis".
A.S. was partially supported by the Russian Foundation for Basic
Research, Grant No. 09-02-12417-ofi-m. This work was started during
his visit to Bologna in 2009 financed by INFN: we thank INFN for
support.


\begin{thebibliography}{10}

\bibitem{BD}
N.~D.~Birrell and P.~C.~W.~Davies, {\em Quantum Fields in Curved
Space} (Cambridge University Press, Cambridge, 1982).

\bibitem{S05}
A.~A.~Starobinsky, JETP Lett. {\bf 82}, 169 (2005).
%%CITATION = ZFPRA,82,187;%%

\bibitem{SYK00}
O.~Seto, J.~Yokoyama and H.~Kodama, Phys.\ Rev.\ D  {\bf 61},
103504 (2000).

\bibitem{FMVV_I}
F.~Finelli, G.~Marozzi, G.~P.~Vacca and G.~Venturi, Phys.\ Rev.\ D
{\bf 65}, 103521 (2002).
%%CITATION = PHRVA,D65,103521;%%

\bibitem{FMVV_II}
F.~Finelli, G.~Marozzi, G.~P.~Vacca, and G.~Venturi,
Phys.\ Rev.\ D {\bf 69}, 123508 (2004).
%%CITATION = PHRVA,D69,123508;%%

\bibitem{FMVV_IV}
F.~Finelli, G.~Marozzi, G.~P.~Vacca and G.~Venturi, Phys.\ Rev.\ D
{\bf 74}, 083522 (2006).
%%CITATION = PHRVA,D74,083522;%%

\bibitem{FMSVV}
F.~Finelli, G.~Marozzi, A.~A.~Starobinsky, G.~P.~Vacca and
G.~Venturi, Phys. Rev. D {\bf 79}, 044007 (2009).
%%CITATION = PHRVA,D79,044007;%%

\bibitem{UM}
Y.~Urakawa and K.-I.~Maeda, Phys. Rev. D {\bf 77}, 024013 (2008);
Phys. Rev. D {\bf 78}, 064004 (2008).
%%CITATION = PHRVA,D77,024013;%%
%%CITATION = PHRVA,D78,064004;%%


\bibitem{JMPW}
T.~M.~Janssen, S.~P.~Miao, T.~Prokopec and R.~P.~Woodard, Class.
Quant. Grav. {\bf 25}, 245013 (2008).
%%CITATION = CQGRD,25,245013;%%



\bibitem{JP}
T.~M.~Janssen and T.~Prokopec, arXiv:0906.0666.
%%CITATION = ARXIV:0906.0666;%%


\bibitem{L82}
A.~D.~Linde, Phys.\  Lett.\  B {\bf 116}, 335 (1982).
%%CITATION = PHLTA,B116,335;%%

\bibitem{S82}
A.~A.~Starobinsky, Phys.\ Lett.\ B {\bf 117}, 175 (1982).
%%CITATION = PHLTA,B117,175;%%

\bibitem{VF82}
A.~Vilenkin and L.~Ford, Phys.\ Rev.\ D {\bf 26}, 1231 (1982).
%%CITATION = PHRVA,D26,1231;%%

\bibitem{S86}
A.~A.~Starobinsky, in {\em Field Theory, Quantum Gravity
and Strings}, eds. H.~J.~De~Vega and N.~Sanchez, Lect.\ Notes\
in\ Physics {\bf 246}, 107 (Springer, New York, 1986).

\bibitem{L86}
A.~D.~Linde, Phys.\ Lett.\ B {\bf 175}, 395 (1986).
%%CITATION = PHLTA,B175,395;%%


\bibitem{LLM94}
A.~D.~Linde, D.~A.~Linde and A.~Mezhlumian, Phys.\ Rev.\  D {\bf
49}, 1783 (1994).
%%CITATION = PHRVA,D49,1783;%%

\bibitem{TW}
N.~C.~Tsamis and R.~P.~Woodard, Nucl.\ Phys.\  B {\bf 724} (2005)
295.
%%CITATION = NUPHA,B724,295;%%


\bibitem{SY}
A.~A.~Starobinsky and J.~Yokoyama, Phys.\ Rev.\ D {\bf 50}, 6357 (1994).
%%CITATION = PHRVA,D50,6357;%%

\bibitem{KO}
E.~O.~Kahya and V.~K.~Onemli, Phys.\ Rev.\ D {\bf 76}, 043512 (2007).
%%CITATION = PHRVA,D76,043512;%%

\bibitem{S78}
A.~A.~Starobinsky, Sov. Astron. Lett. {\bf 4}, 82 (1978).

\bibitem{exponential}
F.~Lucchin and S.~Matarrese, Phys.\ Rev.\ D {\bf 32}, 1316 (1985).
%%CITATION = PHRVA,D32,1316;%%

\bibitem{KST07}
E.~W.~Kolb, A.~A.~Starobinsky and I.~I.~Tkachev, JCAP 0707, 005 (2007).

\bibitem{GR} I.~S.~Gradshteyn and I.~M.~Ryzhik, {\em Table of Integrals,
    Series and Products} (Academic Press, New York, 1980).

\bibitem{Mukhanov} V.~F.~Mukhanov, JETP Lett. {\bf 41}, 493 (1985);
%%CITATION = JTPLA,41,493;%%
Sov.\ Phys.\ JETP {\bf 68} 1297 (1988).
%%CITATION = SPHJA,67,1297;%%

\bibitem{FHLL}
F.~Finelli, J.~Hamann, S.~M.~Leach and J.~Lesgourgues,
arXiv:0912.0522 [astro-ph.CO].
%%CITATION = ARXIV:0912.0522;%%


\bibitem{Marozzi}
G.~Marozzi, Phys.\ Rev.\ D {\bf 76}, 043504 (2007).
%%CITATION = PHRVA,D76,043504;%%

\bibitem{LMY}
M.~Lemoine, J.~Martin and J.~Yokoyama, Phys. Rev. D {\bf 80},
123514 (2009); Europhys. Lett. {\bf 89}, 29001 (2010).
%%CITATION = PHRVA,D80,123514;%%
%%CITATION = EULEE,89,29001;%%


\bibitem{PS92}
D.~Polarski and A.~A.~Starobinsky, Nucl. Phys. B {\bf 385}, 623
(1992).
%%CITATION = NUPHA,B385,623;%%


\end{thebibliography}
\end{document}